\documentclass[amsmath,notitlepage,showkeys,showpacs,superscriptaddress]
{revtex4-1}
\usepackage[utf8]{inputenc}
\usepackage{grffile,graphicx,upgreek,t1enc}

\newcommand{\picref}[1]{Fig.~\ref{fig:#1}}
\newcommand{\eqnref}[1]{Eq.~(\ref{eq:#1})}
\newcommand{\chref}[1]{\ref{ch:#1}}

\usepackage{amsopn}
\DeclareMathOperator{\stdev}{stdev}
\DeclareMathOperator{\Tr}{Tr}

\begin{document}

\title{Towards a continuum model for particle-induced velocity
  fluctuations in suspension flow through a stenosed geometry}

\author{Florian Janoschek}
\email{fjanoschek@tue.nl}
\affiliation{Department of Applied Physics, Eindhoven University of
  Technology, P.\,O. Box 513, 5600\,MB Eindhoven, The Netherlands}
\author{Jens Harting}
\email{j.harting@tue.nl}
\affiliation{Department of Applied Physics, Eindhoven University of
  Technology, P.\,O. Box 513, 5600\,MB Eindhoven, The Netherlands}
\affiliation{Institute for Computational Physics, University of Stuttgart,
  Allmandring 3, 70569 Stuttgart, Germany}
\author{Federico Toschi}
\email{f.toschi@tue.nl}
\affiliation{Department of Applied Physics, Eindhoven University of
  Technology, P.\,O. Box 513, 5600\,MB Eindhoven, The Netherlands}
\affiliation{CNR-IAC, Via dei Taurini 19, 00185 Rome, Italy}

\begin{abstract}
  Non-particulate continuum descriptions allow for computationally
  efficient modeling of suspension flows at scales that are
  inaccessible to more detailed particulate approaches. It is well
  known that the presence of particles influences the effective
  viscosity of a suspension and that this effect has thus to be
  accounted for in macroscopic continuum models. The present paper
  aims at developing a non-particulate model that reproduces not only
  the rheology but also the cell-induced velocity fluctuations,
  responsible for enhanced diffusivity. The results are obtained from
  a coarse-grained blood model based on the lattice Boltzmann
  method. The benchmark system comprises a flow between two parallel
  plates with one of them featuring a smooth obstacle imitating a
  stenosis. Appropriate boundary conditions are developed for the
  particulate model to generate equilibrated cell configurations
  mimicking an infinite channel in front of the stenosis. The averaged
  flow field in the bulk of the channel can be described well by a
  non-particulate simulation with a matched viscosity. We show that
  our proposed phenomenological model is capable to reproduce many
  features of the velocity fluctuations.

\end{abstract}


\maketitle

\section{Introduction}

The presence of particles in a flowing suspension causes
macroscopically relevant effects via small disturbances of the local
velocity of the suspending medium even for laminar homogeneous
flows. This effect can augment the transport through the
suspension~\cite{zydney88,ishikawa08}. An important example is blood
which is a suspension mostly consisting of red blood cells (RBCs) in
blood plasma. Of special interest for medical applications is the
transport of plasma molecules which due to their dimensions can be
assumed to follow the streamlines of the flow and which play a crucial
role in blood clotting phenomena. While advances in the development of
coarse-grained blood models have been made, simulations of realistic
vessel geometries at particulate resolution are still computationally
expensive~\cite{janoschek10,melchionna11}. Therefore, continuous
descriptions of blood are applied frequently to study flows at large
scales, compared to single cells, eventually supported by a continuous
description of the effective transport of cellular and molecular blood
constituents~\cite{boyd05,mikhal12,sorensen99}.

The present work aims instead at developing a continuous description
of the fluid velocity fluctuations and its comparison with large-scale
particulate simulations in an infinite channel with a single
constriction. A similar geometry is modeled, for example, in
Ref.~\cite{ishikawa07}. Defining appropriate boundary conditions for
the inlet and outlet is not trivial in the case of
suspensions. Section~\chref{method} briefly introduces the
coarse-grained particulate blood model~\cite{janoschek10} employed
here. Section~\chref{pi} contains a description of the required
boundary conditions in a parallel implementation. Section~\chref{rec}
deals with the reconstruction of the flow field and of the plasma
velocity fluctuations in a non-particulate simulation. Conclusions are
drawn in Section \chref{conclusions}.

\section{Simulation method}\label{ch:method}

In an earlier work, a simplified particulate blood model was
developed~\cite{janoschek10}. RBCs are described as oblate spheroids
coupled to a lattice Boltzmann (LB) method~\cite{succi01} that
accounts for the blood plasma. One lattice spacing resembles
$0.667\,\upmu\mathrm{m}$, one LB time step
$6.80\times10^{-8}\,\mathrm{s}$. All parameters are chosen as in
Ref.~\cite{janoschek10}. A pair of mutual forces
\begin{equation}\label{eq:deltapppp}
  \mathbf{F}^+
  =
  2n_r^\text{eq}(\bar{\rho},\mathbf{u}{=}\mathbf{0})\mathbf{c}_r
  \text{\quad and\quad}
  \mathbf{F}^-
  =
  -\mathbf{F}^+
\end{equation}
at lattice links connecting two cells corrects for the lack of fluid
pressure at cell-cell
interfaces. $n_r^\mathrm{eq}(\bar{\rho},\mathbf{u})$ stands for the LB
equilibrium distribution function for density $\bar{\rho}$ and
velocity $\mathbf{u}$ of the fluid and $\mathbf{c}_r$ for the
respective lattice vector. The present simulations are the first
application of this model to a situation where---due to the drop of
the pressure $P$ induced by the stenosis---the fluid density $\rho\sim
P$ varies macroscopically across a system with many cells. Under these
conditions, the global average $\bar{\rho}$ in \eqnref{deltapppp} has
to be replaced by a local average $(\bar{\rho}_i+\bar{\rho}_j)/2$
between cells $i$ and $j$ to prevent artificial attraction or
repulsion. $\bar{\rho}_\alpha$ is an average over all fluid sites at
the surface of cell $\alpha$. Further, $\bar{\rho}_i$ is used as the
new fluid density wherever a lattice site occupied by cell $i$ before
is freed and to compute \eqnref{deltapppp} at direct cell-wall
links. Sometimes a slight mass drift is observed whereas a constant
density $\bar{\rho}$ makes significant spatial variations of $\rho$
impossible in the presence of many cells. Therefore, all densities are
rescaled periodically to keep the global average density constant. In
the simulations below, the required rescaling factor differs by less
than $10^{-3}$ from unity if rescaling is performed every $1000$ time
steps.

Since \eqnref{deltapppp} does not account for dissipation at cell-cell
contacts, a further adaption is made: the force on cell $i$ resulting
from a link to a site of cell $j$ is computed not from
\eqnref{deltapppp} but assuming fluid at distribution
$n_r^\mathrm{eq}((\bar{\rho}_i+\bar{\rho}_j)/2,\mathbf{u}_j)$ at the
site belonging to $j$. $\mathbf{u}_j$ is $j$'s local velocity half-way
along the link. The resulting forces on two cells remain opposite but
equal up to first order in velocity. Similarly, for cell-wall links, a
fluid distribution $n_r^\mathrm{eq}(\bar{\rho}_i,\mathbf{0})$ is
assumed at the wall site. At a cell volume fraction of
$\Phi\approx0.43$, the new contact rule leads to an enhancement of the
relative suspension viscosity $\mu_\mathrm{r}$ by $0.5$ to $1$ or
about $10\,\%$ to $30\,\%$ for shear rates
$50\,\mathrm{s}^{-1}<\dot{\gamma}<2\times10^3\,\mathrm{s}^{-1}$. Accurate
sub-lattice corrections for the hydrodynamic interactions of rigid
spheroidal particles near contact have been presented
recently~\cite{janoschek13}.

\section{Periodic inflow boundary conditions}\label{ch:pi}

The transition length that a suspension needs to flow through a
channel until a macroscopically stable state is achieved is known to
be particularly large because of the time required for the particles
to re-distribute in response to the laterally inhomogeneous
flow~\cite{nott94}. Since an appropriately long simulation volume is
computationally too expensive, one could close the geometry in the
flow direction with periodic boundary conditions and simulate for
times long enough that each cell and each volume of fluid passes the
system several times. There is the concrete risk, however, that
because of the same effect~\cite{nott94} the suspension would retain a
memory of the stenosis and the simulation would describe not the flow
through a stenosed channel but, in fact, the flow through an infinite
series of stenoses. Appropriate boundary conditions are required that
provide the cell configurations expected for a long channel without
stenosis. The configuration expected for a long channel, however, is
not known \textit{a priori}. A solution is to develop boundary
conditions that allow to independently simulate a periodic sub-volume
resembling an infinitely long channel while the resulting
configurations are fed into the inlet of the volume that contains the
stenosis. The cells leaving the stenosed sub-volume are then
discarded.

\begin{figure}
  \centering
  \includegraphics[width=0.49\textwidth]
  {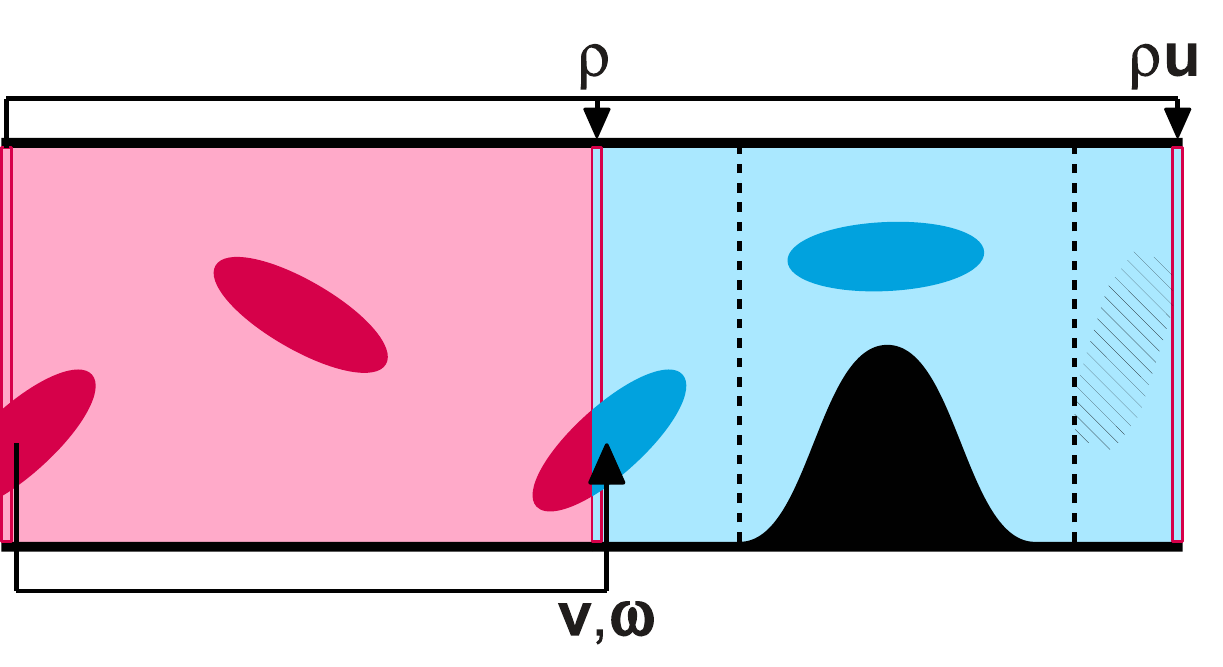}
  \caption{\label{fig:periodic-inflow-4}Schematic outline of the
    boundary conditions: fluid density $\rho$ and mass flow
    $\rho\mathbf{u}$ are copied from the first lattice layer of the
    periodic sub-volume to the first and last layer of the
    non-periodic volume, respectively. Translational and rotational
    cell velocities $\mathbf{v}$ and $\boldsymbol{\omega}$ of original
    periodic cells are imposed on copies in the non-periodic inlet
    region. The dashed lines indicate the end of this region and the
    start of the outflow region where cells are replaced with fluid.}
\end{figure}

The full procedure is sketched in \picref{periodic-inflow-4}. Copies
of model cells in the periodic sub-volume are generated as soon as
they approach the periodic boundary. While the copied cells are
entering the non-periodic volume, their motion is still prescribed by
the motion of the original cells. Once the cells in the non-periodic
sub-volume have reached a longitudinal position about one cell
diameter away from the entrance layer, the connection to the original
cells is dismissed and the copied cells interact with the fluid and
the surrounding cells as free particles. Before a cell in the
non-periodic sub-volume reaches the last lattice layer, the sites
occupied by it are replaced with fluid initialized according to the
local rigid body motion of the cell and the average fluid density
$\bar{\rho}_i$. On-site boundary conditions~\cite{hecht10} impose the
density $\rho$ and the mass flow $\rho\mathbf{u}$ obtained for every
site of the first lattice layer of the periodic sub-volume on the
first and the last layer of the non-periodic sub-volume,
respectively. If, in the case of the mass flow, the source position is
occupied by a model cell, an estimate is made based on its rigid body
motion and the average fluid density.

In a parallel implementation, additional communication in the
longitudinal direction is required with respect to a conventional
three-dimensional domain decomposition scheme with periodically closed
topology. The necessary communication steps are sketched exemplarily
for the case of a two-dimensional decomposition into $5\times2$
computational domains in
\picref{periodic-inflow-parallelization}. First, two-way communication
is needed to establish the link to close the first sub-volume of the
system periodically. Second, the density and mass flow need to be
transferred to the processes holding the first and the last lattice
layer of the remaining non-periodic volume. Similarly, the cell
velocities from the beginning of the periodic sub-volume have to be
sent to the beginning of the non-periodic sub-volume. The last step
might involve communication to more than one destination process if
the end of the entrance region (drawn as dashed line in
\picref{periodic-inflow-parallelization}(b)) lies in another domain
than the sub-volume interface.

\begin{figure}
  \centering
  \includegraphics[height=0.18\columnwidth]
  {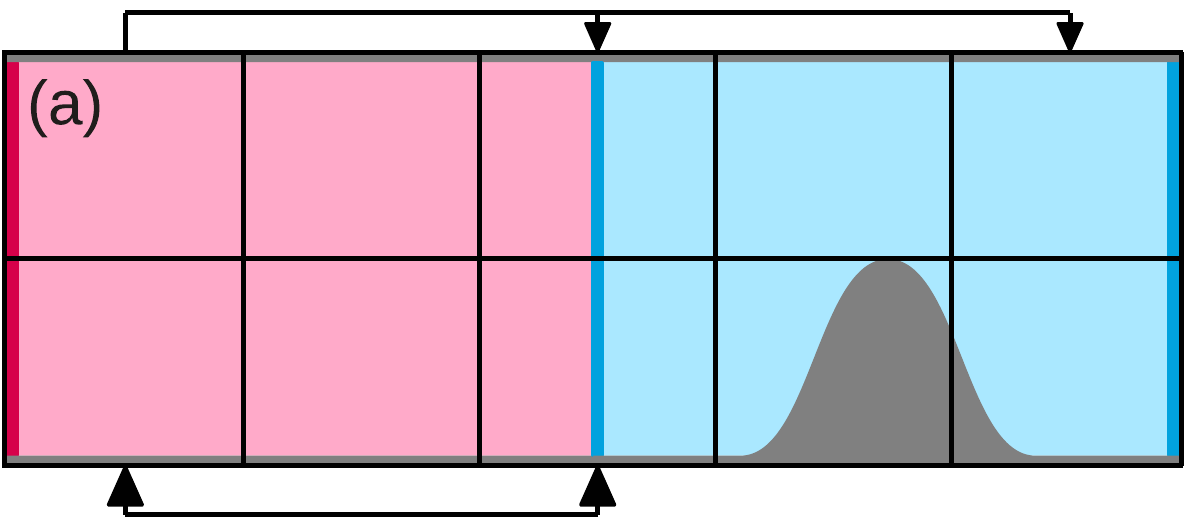}%
  \qquad%
  \includegraphics[height=0.18\columnwidth]
  {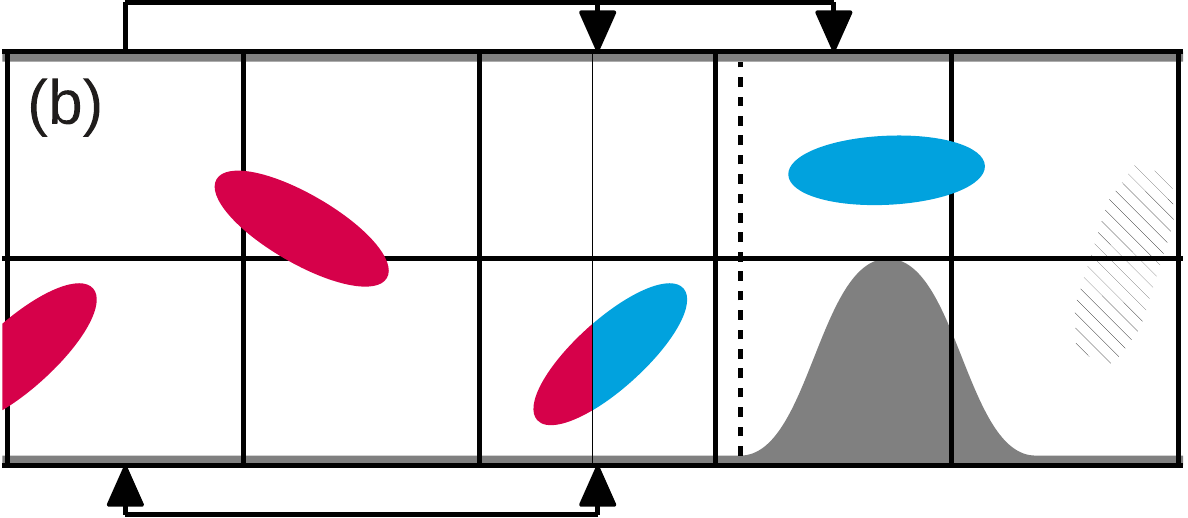}
  \caption{\label{fig:periodic-inflow-parallelization}Communication
    requirements of the boundary conditions. Here an example is shown
    for the case of a decomposition into $5\times2$ domains. Arrows
    indicate the direction along which information is sent. For (a)
    the plasma, the on-site boundary conditions require data from the
    periodic sub-volume which itself needs to be closed
    periodically. For (b) the cells, communication is required for the
    periodic boundaries as well and to prescribe the cell motion in
    the non-periodic entrance region.}
\end{figure}

The simulations are driven by a volume force $g_z$ in $z$-direction
acting on every site within the periodic sub-volume. At sites occupied
by a cell, the force is incorporated by the cell locally. $g_z$ is
updated to a new value $g_z^*=g_zM_z^*/M_z$ at empirically determined
time intervals in order to steer the mass flow $M_z$ towards an
analytically estimated value $M_z^*$ for which the maximum flow
velocity is close to $0.1$ in lattice units as a compromise between
short simulation runs and the stability of the LB method. The
resulting quotient of the maximum velocity $u_\mathrm{max}$ and the
half-width of the opening of the constriction, an estimate for the
average shear rate, amounts to ${\approx}5\times10^4\,\mathrm{s}^{-1}$
in physical units. Though such shear rates, as the maximum flow
velocity $u_\mathrm{max}\approx1\,\mathrm{m}/\mathrm{s}$, under
physiological conditions are not known for vessels with a diameter of
only $85\,\upmu\mathrm{m}$ (see Ref.~\cite{robertson08}), the
geometry can be understood as a very rough simplification of a real
pathological stenosis and its partially curved geometry provides an
interesting and well-defined test case. The Reynolds number is
$\mathrm{Re}=\rho Hu_\mathrm{max}/(\mu_\mathrm{r}\mu_0)\approx9$ with
the plasma viscosity $\mu_0$ when defined using the viscosity of the
blood model for high shear rates as obtained in separate simulations.

\section{Reconstruction of flow velocities and their
  fluctuations}\label{ch:rec}

The following study is performed for flow between two planar walls at
a distance of $85\,\upmu\mathrm{m}$ in the $x$-direction and a length
of $341\,\upmu\mathrm{m}$ in the $z$-direction. In the $y$-direction,
the system is periodic with a depth of $85\,\upmu\mathrm{m}$. The
lower wall carries a sinusoidal ``stenosis'' with a maximum height
$H=42\,\upmu\mathrm{m}$. A mid-link bounce-back scheme ensures no-slip
conditions at all walls. \picref{snapshots} shows an equilibrated
snapshot of both the periodic and the non-periodic sub-volume. The
volume fraction $\Phi=0.4$ leads to a total of ${\sim}10^4$
cells. After the constriction, a cell-free region develops in which
recirculation is found. The flow velocity in $z$-direction $\langle
u_z\rangle$ is reproduced in a continuous simulation without cells but
with a viscosity matched to the high-shear viscosity of the blood
model at $\Phi\approx0.4$ which is achieved with an LB relaxation time
of $\tau=2.15$. The averaging ``$\langle\ldots\rangle$'' is performed
over equilibrated samples at different times and the periodic
$y$-direction.

\begin{figure}
  \centering
  \includegraphics[width=\columnwidth]
  {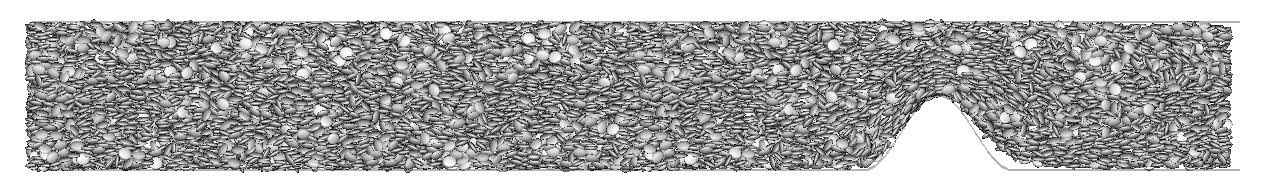}
  \caption{\label{fig:snapshots}Snapshot of both sub-volumes after
    equilibration at a cell volume fraction of $\Phi=0.4$.}
\end{figure}

\picref{continuous-uzf-and-err} compares the velocity fields $\langle
u_z(z,x)\rangle$ from the particulate and $u_z^=(z,x)$ from the
non-particulate simulation. Also the absolute and relative differences
$u_z^-=u_z^=-\langle u_z\rangle$ and $|u_z^-/u_z^+|$ with
$u_z^+=u_z^=+\langle u_z\rangle$ are plotted. Of course, the
continuous model does not take into account the $\Phi$-induced
deviations of $\mu_\mathrm{r}$ close to the solid walls. Thus, $u_z^=$
is too low in the otherwise cell-free boundary layers and consequently
too high in the central region of the channel since the flow rate is
kept constant. Still, the macroscopic features of the velocity field
are reproduced relatively well. The relative difference is less than
$10\,\%$ in the bulk region but higher than $50\,\%$ close to the
walls and particularly in the recirculation zone which has two
reasons: the low velocities there, which make small absolute
differences more visible, and cell-depletion effects.

\begin{figure}
  \centering
  \includegraphics[width=0.49\textwidth]
  {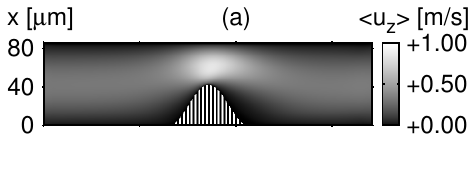}%
  \hfill%
  \includegraphics[width=0.49\textwidth]
  {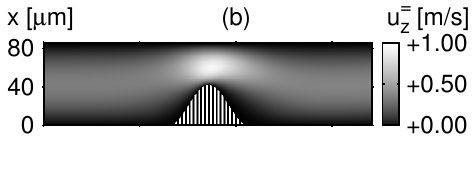}\\[-4ex]
  \includegraphics[width=0.49\textwidth]
  {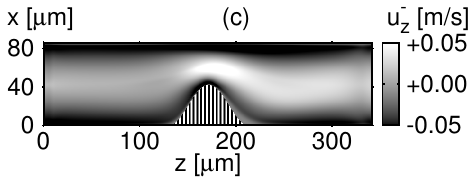}%
  \hfill%
  \includegraphics[width=0.49\textwidth]
  {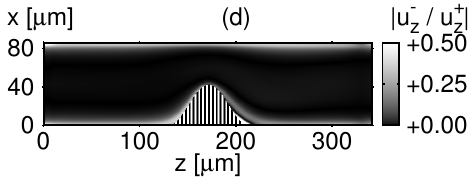}
  \caption{\label{fig:continuous-uzf-and-err}Comparison of (a)
    $\langle u_z\rangle$ at $\Phi=0.4$ and (b) $u_z^=$ at $\Phi=0$
    with matched viscosity. (c) and (d) show the absolute and relative
    differences.}
\end{figure}

Fluctuations of the flow velocity $\mathbf{u}$ are quantified by the
standard deviation
\begin{equation}
  V
  =
  \stdev\,\mathbf{u}
  =
  \left[
    \sum_{\alpha=x,y,z}
    \left(
      \left\langle u_\alpha^2\right\rangle
      -\left\langle u_\alpha\right\rangle^2
    \right)
  \right]^{1/2}
  \text{ .}
\end{equation}
The lateral profile $V(x)$ distant from the constriction is plotted in
\picref{model}. $V$ reaches a maximum about one cell diameter away
from the wall but does not increase further for shorter
distances. Directly at the wall, no fluid is present and $V$ is
zero. In the center of the channel, $V$ approximates a constant value
of about one third of its maximum. In between, a roughly linear
increase towards the wall is observed.

\begin{figure}
  \centering
  \includegraphics[width=0.6\textwidth]
  {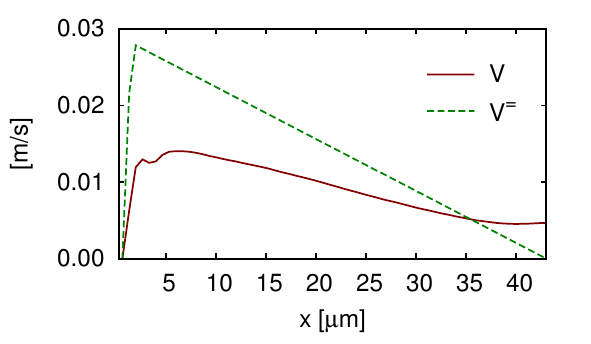}
  \caption{\label{fig:model}Lateral profile of the fluid velocity
    fluctuations $V$ without constriction. Shown is half of the
    channel cross section with the center at the right. Simulation
    data is compared to the dimensional estimate $V^=$ defined in
    \eqnref{estimate} based on the shear rate in a continuous
    simulation.}
\end{figure}

Assuming that in the bulk of the suspension the time scale relevant
for $V$ is the inverse undisturbed shear rate $\dot{\gamma}^{-1}$,
which at scales much larger than single cells can easily be obtained
from a non-particulate simulation, and that the relevant length scale
must be of the order of the size of the cells, a dimensional estimate
\begin{equation}\label{eq:estimate}
  V^=
  \simeq
  \dot{\gamma}R_\parallel
\end{equation}
for $V$ induced by cells with shorter half-axes $R_\parallel$ is
made. Also this estimate is plotted in \picref{model}, where the shear
rate is obtained numerically from the continuous simulation as the
derivative $\dot{\gamma}=\partial_xu_z^=$. For a Poiseuille flow,
$\dot{\gamma}$ is a linear function of the lateral position and
reaches its maximum at the wall. The decrease directly at the wall
visible in \picref{model} is caused by the absence of fluid at the
wall itself and by artefacts of the numerical
derivation. \eqnref{estimate} reproduces the order of magnitude of the
plasma velocity fluctuations correctly with an over-prediction of less
than a factor two in the largest part of the plot. Also the linear
dependence on the lateral position is observed for $V$ as well, at
least in parts of the curve. The two regions where deviations from the
linear shape are visible have a width of about one cell diameter and
can indeed be attributed to cell effects: since cells cannot pass the
vessel wall, the cell concentration immediately near to the wall is
reduced and the motion of cells close to the wall is hindered. Both
effects reduce cell-induced velocity fluctuations. On the other hand,
even the plasma at the channel center, where the averaged shear rate
vanishes, is disturbed by nearby cells that at their off-center
position experience non-zero shear.

At last the applicability of the model to the full geometry with the
stenosis is demonstrated. This requires the scalar shear rate to be
determined in an isotropic way as
$\dot{\gamma}=\sqrt{2D_\mathrm{II}}$, similarly as in
Ref.~\cite{boyd07} based on the second invariant $D_\mathrm{II}=\Tr
S_{\alpha\beta}^2-\Tr^2 S_{\alpha\beta}$ of the numerically computed
strain rate tensor $S_{\alpha\beta}=(\partial_\beta
u_\alpha^=+\partial_\alpha u_\beta^=)/2$ assuming incompressibility
$\Tr S_{\alpha\beta}=0$ where both $\alpha$ and $\beta$ stand for $x$
and $z$. For the whole geometry, \picref{continuous-stddu} compares
the resulting $V^=$ to $V$. While, as already in \picref{model}, the
agreement is not perfect, many features of $V$ are reproduced
qualitatively and it has to be pointed out that the correct order of
magnitude is well captured without fitting parameters. As expected
from \picref{model}, the absolute deviations are strongest at the
walls and at the center of the channel, especially at the constriction
where $\dot{\gamma}$ is highest. Still, the relative errors are below
$50\,\%$ in large parts of the geometry.

\begin{figure}
  \centering
  \includegraphics[width=0.49\textwidth]
  {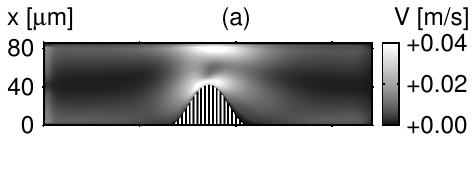}%
  \hfill%
  \includegraphics[width=0.49\textwidth]
  {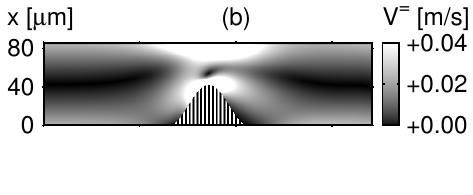}\\[-4ex]
  \includegraphics[width=0.49\textwidth]
  {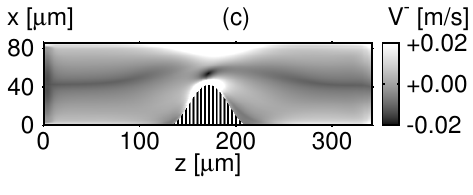}%
  \hfill%
  \includegraphics[width=0.49\textwidth]
  {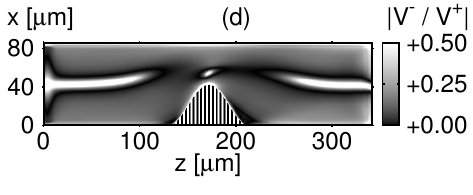}
  \caption{\label{fig:continuous-stddu}Comparison of (a) $V$ as
    obtained from a particulate simulation at $\Phi=0.4$ and (b) the
    reconstruction $V^=$ defined in \eqnref{estimate} based on the
    isotropic shear rate obtained from a non-particulate simulation
    ($\Phi=0$) with matched viscosity. (c) and (d) show the absolute
    and relative differences $V^-=V^=-V$ and $|V^-/V^+|$ with
    $V^+=V^=+V$.}
\end{figure}

\section{Conclusions}\label{ch:conclusions}

Boundary conditions suitable for efficient modeling of equilibrated
suspension flow through an infinite channel followed by a single
stenosis were implemented in a parallel LB code. The time-averaged
flow field produced by a coarse-grained particulate blood model was
reproduced by a non-particulate model with a matched
viscosity. Moreover, the cell-induced plasma velocity fluctuations in
the bulk of the suspension can be well understood from a simple
dimensional argument and therefore may be reproduced qualitatively
from the non-particulate simulation.

A continuous model for blood can be run at reduced resolution and
therefore allows to simulate flows in larger geometries more
efficiently, stretching to arteries or veins~\cite{boyd05,mikhal12}.
The estimate \eqnref{estimate} would allow to equip such models with a
qualitative prediction of the RBC-induced plasma velocity
fluctuations. The combined model might be useful for studying
transport phenomena in geometries inaccessible to the more expensive
particulate blood models. In view of these applications the deviations
of the continuous model at cell-depleted boundary layers are no
serious limitation since these scales would not be resolved in a
continuous simulation at a practical spatial resolution. A next
important step would be to connect the velocity fluctuations to an
effective diffusivity for scalar transport in the medium.

\section*{Acknowledgments}

The authors acknowledge fruitful discussions with Francesca Storti,
Frans van de Vosse, Simone Melchionna, and Sauro Succi, financial
support from the TU/e High Potential Research Program, and computing
resources from the J\"ulich Supercomputing Centre granted through PRACE.

\end{document}